\documentstyle[12pt]{article}
\def \bea{\begin{eqnarray}}
\def \beq{\begin{equation}}

\def \eea{\end{eqnarray}}
\def \eeq{\end{equation}}
\def \gsim{\stackrel{>}{\sim}}

\def \ket#1{| #1 \rangle}
\def \mat#1#2{\langle #1 | #2 \rangle}
\def \od{\overline{D}^0}
\def \pp{\psi'}
\def \ppp{\psi''}

\begin{document}
\Large
\centerline {\bf Charmless final states and S--D-wave mixing}
\centerline{\bf in the $\ppp$
\footnote{Enrico Fermi Institute preprint EFI 01-21, hep-ph/0105327.
Submitted to Physical Review D.}}
\normalsize
\bigskip
 
\centerline{Jonathan L. Rosner~\footnote{rosner@hep.uchicago.edu}}
\centerline {\it Enrico Fermi Institute and Department of Physics}
\centerline{\it University of Chicago, 5640 S. Ellis Avenue, Chicago, IL 60637}
\medskip
\centerline{(Received June 2001)}
\medskip
 
\begin{quote}

The $\ppp = \psi(3770)$ resonance is expected to be mainly $c \bar c(1^3D_1)$,
but tensor forces and coupling to charmed particle pairs can mix it with
$\pp(2^3S_1)$ and other states.  Implications of this mixing for decays of
$\ppp$ to non-charmed final states are discussed.  (i) The ratio $\Gamma(\ppp
\to \gamma + \chi_{c2})/ \Gamma(\ppp \to \gamma + \chi_{c0})$ is expected to
be highly suppressed if $\ppp$ is a pure D-wave state, and is enchanced by
mixing.  (ii) The expected decay $\pp \to \rho \pi$ and other ``missing'' modes
can appear as corresponding $\ppp$ partial widths, enhanced by a factor
depending on the mixing angle.  General arguments then suggest a branching
ratio of about 1\%, give or take a factor of 2, for charmless hadronic
decays of $\ppp$.  (iii)  Enhancements can appear in penguin
amplitudes in $B$ decays, $B \to K \eta'$ branching ratios, and direct
CP-violating asymmetries in $B \to K \pi$ decays.

\end{quote}
\medskip

\centerline{PACS numbers:  13.25.Gv, 13.20.Gd, 14.40.Gx, 12.39.Jh}

\section{Introduction}

The lowest resonance in electron-positron collisions above charmed particle
pair production threshold is the $\ppp = \psi(3770)$, discovered somewhat
after the $J/\psi(3097)$ and the $\pp = \psi(3686)$ \cite{Rap}.\footnote{The
numbers in parentheses indicate the masses of the particles, in MeV/$c^2$.}
It provides a rich source of $D^0 \od$ and $D^+ D^-$ pairs, as
anticipated theoretically \cite{Eetal}.  The largest data
sample of $\ppp$ decays studied so far, by the Mark III Collaboration at the 
Stanford electron-positron collider SPEAR \cite{MkIII}, has been $9.56 \pm
0.48$ pb$^{-1}$.  Plans are under way to accumulate as much as 3 fb$^{-1}$
at the Cornell Electron Storage Ring (CESR), which will permit much more
incisive tests of a number of open questions \cite{Wkshp}.  In the present
note we discuss several of these which involve observation of
{\it non-charmed final states} of the $\ppp$.  These have been studied in two
previous Ph.~D. theses \cite{Zhu,Walid} based on the Mark III data.

The $\ppp$ is the only present candidate for a D-wave $(l = 2)$
quarkonium level. (Strategies for finding the corresponding $b \bar b$ levels
have been noted in Refs.\ \cite{KR,GRD}.) Although it is primarily $c \bar
c(1^3D_1)$, \footnote{We shall use spectroscopic notation $n^{2S+1}L_J$, where
$n = 1, 2, 3, \ldots$ is the radial quantum number; $S = 0$ or 1 is the
$Q \bar Q$ spin; $L = S,~P,~D, \ldots$ ($l = 0, 1, 2, \ldots$) is the orbital
angular momentum; and $J = 0, 1, 2, \ldots$ is the total spin.} its leptonic
width (quoted in Table I \cite{MkIII,PDG}) indicates a contribution from mixing
with S-wave states, such as the nearby $\pp(2^3S_1)$ and to a lesser extent
with $J/\psi(1^3S_1)$ \cite{Rich} and $n \ge 3$ S-wave states above 4
GeV/$c^2$.  Early calculations of this mixing based on contributions from
intermediate real and virtual states of charmed particle pairs \cite{Eetal}
predicted a $\ppp$ contribution to the $e^+ e^- \to D \bar D$ cross section
which indicated the utility of this state as a ``charm factory'' and
predicted its leptonic width quite well.\footnote{For later discussions of
mixing due to coupled-channel effects see \cite{ECC}.}  It was later found that
mixing due to a tensor force based on perturbative QCD also was adequate to
explain the observed leptonic width \cite{MR}.  Probably both perturbative
and non-perturbative (e.g., coupled-channel) effects are present.

\begin{table}
\caption{Properties of the $\ppp = \psi(3770)$}
\begin{center}
\begin{tabular}{c c c c c} \hline
Mass (MeV/$c^2$) & $\Gamma_{\rm tot}$ (MeV) & $\Gamma_{ee}$ (keV) &
${\cal B}(D^0 \od)$ & ${\cal B}(D^+ D^-)$ \\ \hline
$3769.9 \pm 2.5$ & $23.6 \pm 2.7$ & $0.26 \pm 0.04$ & 58\% & 42\% \\ \hline
\end{tabular}
\end{center}
\end{table}

The mixing of the $\ppp$ with other states can affect both its decays
and those of the other states.  In Section II we discuss a simplified model
for $\pp$--$\ppp$ mixing and its implications for leptonic and radiative
partial decay rates of these states.  The ratio $\Gamma(\ppp \to \gamma +
\chi_{c2})/ \Gamma(\ppp \to \gamma + \chi_{c0})$ is expected to be highly
suppressed if $\ppp$ is a pure D-wave state, but could be enhanced by mixing
\cite{Zhu,KR,YNY,KL,B88}.

The ``missing decay modes'' of the $\pp$ \cite{psip}, such as $\rho \pi$ and
$K^* \bar K + {\rm~c.c.}$, are a long-standing puzzle
\cite{rhopi,Suz,fsi,GW,FK}.
Recently Suzuki \cite{Suz01} showed that if a $\pp$ decay amplitude due
to coupling to virtual (but nearly on-shell) charmed particle pairs interferes
destructively with the standard three-gluon amplitude, the suppression of
these (and other) modes in $\pp$ final states can be understood.  We pursue
this suggestion further in Section III using the $\pp$--$\ppp$ mixing
model described earlier.  We propose that as a result of coupled-channel
effects the expected decay width $\Gamma(\pp \to \rho \pi) \simeq 0.5$ keV and
other ``missing'' modes could show up as corresponding partial widths in $\ppp$
decays, possibly enhanced by a considerable factor depending on the mixing
angle.  Since the latter state has a total width nearly 100 times that of the
$\pp$, each of these partial widths still corresponds to a small branching
ratio.

If coupling to charmed particle pairs is responsible for mixing the $\pp$
and the $\ppp$, and for significant effects on non-charmed final states
in decays of both particles, it is likely that virtual or real $D^{(*)}
\bar D^{(*)}$ pairs produced in low partial waves in other contexts may
undergo significant rescattering into non-charmed final states.  Foremost
among these cases are the decays of $B$ mesons, which can involve such pairs
via the subprocesses $\bar b \to \bar c c \bar s$ or $\bar b \to \bar c c 
\bar d$.  The re-annihilation of the final $c \bar c$ pair can lead to an
effective $\bar b \to \bar s$ or $\bar b \to \bar d$ penguin amplitude
\cite{fsi,Dun,Ciu,KLS}, which appears to be needed in understanding large
branching ratios for $B \to K \eta'$ \cite{eta} and $B \to K \pi$.  Moreover,
Suzuki \cite{Suz01} has proposed that this reannihilation, at least in
$\pp$ decays, is associated with a large final-state phase. We discuss
implications of this suggestion for CP violation in $B$ decays in Section IV,
while Section V concludes.

\section{Radiative $\ppp$ decays}

The relative branching ratios for radiative decays to $\chi_c$ ($1^3P_1$)
states are very different for $2S$ and $1D$ states.  The observation
of radiative decays $\ppp \to \gamma + \chi_c$ can determine the degree to
which the $\ppp$ is mixed with an S-wave state \cite{Zhu,KR,YNY,KL,B88}.

The rates for electric dipole ($E1$) transitions in quarkonium can be written
\beq \label{eqn:rate}
\Gamma = \frac{4}{3} e_Q^2 \alpha \omega^3 C \langle r \rangle^2~~~,
\eeq
where $e_Q$ is the quark charge (in units of $|e|$), $\alpha = 1/137.036$ is
the fine-structure constant, $\omega$ is the photon energy, and $\langle r
\rangle$ is the matrix element of $r$ between initial and final radial
wave functions.  The coefficients $C$ are summarized in Table II, where
we compare relative rates for $E1$ transitions from $\ppp$ to $\chi_c$
states under the two extreme assumptions of a pure S-wave or a pure D-wave.
The distinctive pattern associated with the pure $^3D_1$ configuration is
a ratio ${\cal B}(\gamma + \chi_{c1})/{\cal B}(\gamma + \chi_{c0}) = 0.3$
and an almost complete suppression of the ratio ${\cal B}(\gamma + \chi_{c2})
/{\cal B}(\gamma + \chi_{c0})$.

\begin{table}
\caption{Comparison of transitions $\ppp \to \gamma \chi_c$ under the
assumptions of a pure S-wave or D-wave initial state.  Coefficients $C$ are
those in the expression (\ref{eqn:rate}) for electric dipole transitions.}
\begin{center}
\begin{tabular}{c c c c c c c} \hline
Final   & $\omega$ & \multicolumn{2}{c}{Pure $^3S_1$}
                          & & \multicolumn{2}{c}{Pure $^3D_1$} \\
\cline{3-4} \cline{6-7}
state   &  (MeV)   & $C$ & $\Gamma(^3P_J)/\Gamma(^3P_0)$
                          & & $C$ & $\Gamma(^3P_J)/\Gamma(^3P_0)$ \\ \hline
$^3P_0$ &    338   & 1/9 &  1 & &  2/9 &   1   \\
$^3P_1$ &    250   & 1/3 & 1.22 & & 1/6 &  0.30 \\
$^3P_2$ &    208   & 5/9 & 1.16 & & 1/90 & 0.012 \\ \hline
\end{tabular}
\end{center}
\end{table}

A more detailed model can be constructed by assuming that the $\ppp$
is a mixture of a $1^3D_1$ and a $2^3S_1$ state \cite{B88}:
\beq \label{eqn:mix}
\ket{\ppp} =   \ket{1^3D_1} \cos \phi + \ket{2^3S_1} \sin \phi~~,~~~
\ket{\pp}  = - \ket{1^3D_1} \sin \phi + \ket{2^3S_1} \cos \phi~~~.
\eeq
The leptonic widths of $\ppp$ and $\pp$ are then \cite{Nov}
\beq
\Gamma(\ppp \to e^+ e^-) = \frac{4 \alpha^2 e_c^2}{M_{\ppp}^2} \left|
\sin \phi R_{2S}(0) + \frac{5}{2 \sqrt{2} m_c^2} \cos \phi {R''}_{1D}(0)
 \right|^2~~,
\eeq
\beq
\Gamma(\pp \to e^+ e^-) = \frac{4 \alpha^2 e_c^2}{M_{\pp}^2} \left|
\cos \phi R_{2S}(0) - \frac{5}{2 \sqrt{2} m_c^2} \sin \phi {R''}_{1D}(0) 
 \right|^2~~,
\eeq
where $e_c = 2/3$, $R_{2S}(0) = (4 \pi)^{1/2} \Psi_{2S}(0)$ is the radial
$2S$ wave function at $r=0$, and ${R''}_{1D}(0)$ is the second derivative
of the radial $2D$ wave function at the origin.  The values $R_{2S}(0) =
0.734$ GeV$^{3/2}$ and $5R_{1D}''(0)/(2 \sqrt{2}m_c^2) = 0.095$ GeV$^{3/2}$
were taken in Ref.\ \cite{B88}.  Assuming a common QCD correction to
$\pp$ and $\ppp$ leptonic widths, we then fit the ratio
\beq
\frac{M_{\ppp}^2 \Gamma(\ppp \to e^+ e^-)}{M_{\pp}^2 \Gamma(\pp \to e^+ e^-)}
= \left| \frac{0.734 \sin \phi + 0.095 \cos \phi}
              {0.734 \cos \phi - 0.095 \sin \phi} \right|^2
= 0.128 \pm 0.023~~,
\eeq
with solutions $\phi = (12 \pm 2)^{\circ}$ or $\phi = -(27 \pm 2)^{\circ}$.
These values agree with those of Kuang and Yan \cite{KY90}, whose $\theta$
is the same as our $- \phi$.  As they note, the smaller-$|\phi|$ solution
is consistent with coupled-channel estimates \cite{EGKLY,CC} and with the
ratio of $\pp$ and $\ppp$ partial widths to $J/\psi \pi \pi$.

A nonrelativistic calculation along the lines of Ref.\ \cite{YNY}
then yields the following predictions \cite{B88}:
\beq
\Gamma(\ppp \to \gamma \chi_{c0}) = 145~{\rm keV} \cos^2 \phi
 (1.73 + \tan \phi)^2~~,
\eeq
\beq
\Gamma(\ppp \to \gamma \chi_{c1}) = 176~{\rm keV} \cos^2 \phi
 (-0.87 + \tan \phi)^2~~,
\eeq
\beq
\Gamma(\ppp \to \gamma \chi_{c2}) = 167~{\rm keV} \cos^2 \phi
 (0.17 + \tan \phi)^2~~,
\eeq
\beq
\Gamma(\pp \to \gamma \chi_{c0}) = 67~{\rm keV} \cos^2 \phi
 (1 - 1.73 \tan \phi)^2~~,
\eeq
\beq
\Gamma(\pp \to \gamma \chi_{c1}) = 56~{\rm keV} \cos^2 \phi
 (1 + 0.87 \tan \phi)^2~~,
\eeq
\beq
\Gamma(\pp \to \gamma \chi_{c2}) = 39~{\rm keV} \cos^2 \phi
 (1 - 0.17 \tan \phi)^2~~.
\eeq
Other predictions are given, for example, in Ref.\ cite{GZS}.  Zhu
has apparently neglected to take account of relative signs of S-wave and
D-wave contributions in the first three of the above equations when presenting
his results for mixed states (Fig.\ 1.6.2, Ref.\ \cite{Zhu}).
For small $\phi$, as suggested by the $\pp$ and $\ppp$ leptonic widths, the
experimental rates for the $\pp$ radiative decays are about a factor
of three below these predictions \cite{PDG}, probably as a result of
relativistic corrections \cite{MR,MB}.  The $\pp$ decays are expected to be
particularly sensitive to such corrections as a result of the node in the
$2S$ wave function; it is possible that the $\ppp$ predictions could be more
reliable, since neither the $1D$ nor $1P$ radial wave functions has a node.

Results for $\ppp$ radiative decays \cite{Zhu}, for $\sigma(e^+ e^-
\to \ppp) \equiv \sigma(\ppp) = 5.0 \pm 0.5$ nb, are:
\beq
\Gamma(\ppp \to \gamma \chi_{c0}) = 510 \pm 190~{\rm keV}~~,
\eeq
\beq
\Gamma(\ppp \to \gamma \chi_{c1}) = 440 \pm 160~{\rm keV}~~,
\eeq
\beq
\Gamma(\ppp \to \gamma \chi_{c2}) \le 520~{\rm keV}~(90\% {\rm~c.l.})~~.
\eeq
These partial widths scale as $1/\sigma(\ppp)$.
So far it does not seem possible to reconcile the central values of these
results with the values of $\phi$ suggested earlier.\footnote{The solution with
$\phi = 12^{\circ}$, favored by coupled-channel calculations \cite{EGKLY,CC},
predicts $\Gamma(\ppp \to \gamma \chi_{c(0,1,2)}) = (524,~73,~61)$ keV,
implying that the $\chi_{c1}$ signal of Ref.\ \cite{Zhu} should not be
confirmed.}  The model for mixing
between $\pp$ and $\ppp$ may be oversimplified, and relativistic corrections
undoubtedly play a role.  Nevertheless, the above results bear revisiting with
improved statistics.  The search for a 338 MeV monochromatic photon in the
decays of the $\ppp$ would represent a worthwhile first step in the
determination of this interesting resonance's mixing parameters.
 
\section{Missing modes of the $\pp$}

F. A. Harris \cite{FH} has summarized a wide class of hadronic decay modes of
the $\pp$ which appear to be suppressed relative to expectations.  Of these
the foremost is the $\rho \pi$ final state, with $K^+ K^{*-}(892)
+ {\rm c.c.}$ in second place.  Let us review the expectations and the data for
these two modes.  (The decay $\pp \to K^0 \overline{K}^{*0}(892) + {\rm c.c.}$
has been observed with a branching ratio of $(8.1 \pm 2.4 \pm 1.6) \times
10^{-5}$ which indicates the contribution of a significant one-virtual-photon
contribution \cite{Suz,fsi,Suz01}, and we shall not discuss it further.)

We summarize in Table III the total widths, branching ratios, and derived
partial widths for $J/\psi$ and $\pp$ decays into $\rho \pi$ and $K^+
\overline{K}^*(892)^-$, as well as the partial widths predicted for the
$\pp$ decays to these final states.  Both hadronic and leptonic decay
rates are proportional to the square of the wave function at the origin
$|\Psi(0)|^2$.  Although one might expect an additional factor of $1/M_V^2$,
where $M_V$ is the mass of the decaying vector meson, entering into the
leptonic width, we shall ignore this effect, since it is probably offset by
a (form) factor suppressing the hadronic decay of the higher-mass $\pp$ into 
low-multiplicity final states such as $\rho \pi$.  Then
we expect for any hadronic final state $f$ \cite{rhopi,Suz01,FH}
\beq
\Gamma(\pp \to f) = \Gamma(J/\psi \to f) \frac{\Gamma_{ee}(\pp)}
{\Gamma_{ee}(J/\psi)}~~.
\eeq
This relation has been used to predict the quantities ${\Gamma_{\rm pred}}$
in Table III.  One sees that $\pp \to \rho \pi$ is suppressed by a factor of
at least $\sim 50$ with respect to na\"{\i}ve expectations, while the
corresponding factor for $K^+ K^{*0}(892) + {\rm c.c.}$ is at least $\sim 20$.  

\begin{table}
\caption{Total widths, branching ratios, and derived partial widths for
$J/\psi$ and $\pp$ decays.}
\begin{center}
\begin{tabular}{c c c c c c c} \hline
Decay mode & \multicolumn{2}{c}{$J/\psi$ decays \cite{PDG}} &
  & \multicolumn{3}{c}{$\pp$ decays \cite{FH}} \\
  & \multicolumn{2}{c}{$\Gamma_{\rm tot} = 87 \pm 5~{\rm keV}$} &
  & \multicolumn{3}{c}{$\Gamma_{\rm tot} = 277 \pm 31~{\rm keV}$ \cite{PDG}} \\
  & \multicolumn{2}{c}{$\Gamma_{ee} = 5.26 \pm 0.37~{\rm keV}$} &
  & \multicolumn{3}{c}{$\Gamma_{ee} = 2.12 \pm 0.18~{\rm keV}$ \cite{PDG}}
\\ \cline{2-3} \cline{5-7}
  & ${\cal B}$ & $\Gamma$ (keV) & & ${\cal B}$ & $\Gamma$ (eV)
  & ${\Gamma_{\rm pred}}^{~a}$ (eV) \\ \hline
$\rho \pi$ & $(1.27 \pm 0.09)\%$ & $1.10 \pm 0.10$ &
  & $< 2.8 \times 10^{-5}$ & $ < 8.6$ & $443 \pm 63$ \\
${K^+ K^{*-}(892)}^b$ & $(0.50 \pm 0.04)\%$ & $0.44 \pm 0.04$ &
  & $< 3.0 \times 10^{-5}$ & $ < 9.2$ & $177 \pm 24$ \\ \hline
\end{tabular}
\end{center}
\leftline{$^a$ Based on prescription given in text. $^b$ Plus c.c.}
\end{table}

Suzuki \cite{Suz01} has proposed that the coupling of $\pp$ to virtual
pairs of charmed particles could provide an amplitude which interferes
destructively with the perturbative QCD process $\pp \to 3g$ in the specific
cases of $\rho \pi$ and $K \overline{K}^*(892) + {\rm c.c.}$ hadronic decays.
If this is the case, and if virtual charmed particle pairs also play a role
in mixing $\pp$ and $\ppp$, we would expect a similar amplitude to
contribute to $\ppp \to D^{(*)} \overline{D}^{(*)} \to \rho \pi$ or $K
\overline{K}^*(892) + {\rm c.c.}$

In the absence of a detailed coupled-channel analysis, let us assume that
the main effect on $\pp$ and $\ppp$ of their mutual coupling to charmed
particle pairs is precisely the mixing discussed in the previous section.
Let us assume that this mixing and the couplings of $\pp$ and $\ppp$ to
$\rho \pi$ and $K \overline{K}^*(892) + {\rm c.c.}$ are such as to
cancel the $\pp$ hadronic widths to these final states [which are related to
one another by flavor SU(3)].  In this case we have
$$
\mat{\rho \pi}{\pp} =
\mat{\rho \pi}{2^3S_1} \cos \phi - \mat{\rho \pi}{1^3D_1} \sin \phi = 0~~,
$$
\beq
\mat{\rho \pi}{\ppp} =
\mat{\rho \pi}{2^3S_1} \sin \phi + \mat{\rho \pi}{1^3D_1} \cos \phi =
\mat{\rho \pi}{2^3S_1}/ \sin \phi~~,
\eeq
so that {\it the missing $\rho \pi$ (and related) decay modes of $\pp$ show up
instead as decay modes of $\ppp$, enhanced by the factor of $1/\sin^2 \phi$}.
The possible effects of this enhancement are shown in Table IV for the two
solutions for $\phi$.  One expects ${\cal B}(\ppp \to \rho \pi) \simeq 10^{-4}$
for $\phi \simeq -27^\circ$ and $\simeq 4 \times 10^{-4}$ for the favored value
$\phi \simeq 12^\circ$.  Either branching ratio is compatible
with the current upper bound ${\cal B}(\ppp \to \rho \pi) < 1.3 \times
10^{-3} \times[5~{\rm nb}/\sigma(\ppp)]$ \cite{Zhu}.

\begin{table}
\caption{Predicted $\ppp \to \rho \pi$ partial widths and branching ratios
for two solutions of mixing angle $\phi$.}
\begin{center}
\begin{tabular}{c c c} \hline
$\phi$ ($^\circ$) & $- 27 \pm 2$ & $12 \pm 2$ \\
$1/\sin^2 \phi$ & $4.8 \pm 0.6$ & $22 \pm 6$ \\
$\Gamma(\ppp \to \rho \pi)$ (keV) & $2.1 \pm 0.4$ & $9.8 \pm 3.0$ \\
${\cal B}(\ppp \to \rho \pi)~(10^{-4})$ & $0.9 \pm 0.2$ & $4.1 \pm 1.4$ \\
\hline
\end{tabular}
\end{center}
\end{table}

An alternative mechanism discussed by Suzuki \cite{Suz01} for introducing
an additional non-perturbative $\pp$ decay amplitude is mixing with a
vector glueball state (first discussed in the context of $J/\psi$ decays
\cite{glu}).  In this case the $\ppp$ is permitted, but not required, to mix
with the vector glueball, so there is no particular reason for the missing
partial widths for $\pp$ decays to show up as corresponding $\ppp$ partial
decay rates.

G\'erard and Weyers \cite{GW} have proposed that the three-gluon decay of the
$\pp$ is absent or suppressed, and that the $\pp$ decays to
hadrons instead mainly via a two-step process involving an intermediate
$c \bar c(^1P_1)$ state.  Feldmann and Kroll \cite{FK} have proposed
that the $J/\psi \to \rho \pi$ decay is {\it enhanced} (rather than $\pp \to
\rho \pi$ being suppressed) by mixing of the $J/\psi$ with light-quark states,
notably $\omega$ and $\phi$.  Both mechanisms do not imply any special role for
$\ppp$ charmless decays.  Arguments against them raised in the last of Refs.\
\cite{rhopi} and in Ref.\ \cite{FH} include the appearance of certain
unsuppressed light-quark decay modes of the $\pp$ and the lack of evidence for
helicity suppression in $J/\psi$ decays involving a single virtual photon.

As Suzuki has noted, the cases of suppressed hadronic final states of the
$\pp$ cannot extend to all its decays; indeed, the total hadronic width
of $\pp$ exceeds estimates based on extrapolating from the $J/\psi$ using
perturbative QCD by some 60--70\% \cite{Suz01,GL}.  The non-perturbative effect
of coupling to virtual charmed particle pairs, followed by the re-annihilation
of these pairs into non-charmed final states, must thus be responsible for
some tens of keV of the total width of the $\pp$ in Suzuki's scheme.  

A corresponding effect in the decays of the $\ppp$, which is about 85
times as wide as the $\pp$, would contribute at most a percent to its
total width.  Present searches for non-charmed decays of the $\ppp$
\cite{Zhu,Walid} are not sensitive enough to exclude this possibility since
they did not compare on-resonance data with data taken off-resonance at
a sufficiently close energy \cite{JTpc}.

A related method allows one to estimate the partial decay rate of $\ppp$ to
non-charmed final states.  The branching ratio ${\cal B}(J/\psi \to \rho \pi)$
is $(1.27 \pm 0.09)\%$.  Since about 1/3 of $J/\psi$ decays can be ascribed to
non-$3 g$ mechanisms, we expect $\rho \pi$ to account for about 2\% of all {\it
hadronic} $J/\psi$ decays, and thus no more than this percentage of $\ppp$
hadronic charmless decays. (The availability of more final states undoubtedly
reduces the $\rho \pi$ fraction in comparison with $J/\psi$ hadronic decays.)
We thus estimate for hadronic charmless decays ${\cal B}(\ppp) \gsim 2 \times
10^{-4} /2\% \simeq 1\%$, again give or take a factor of 2 depending on the
sign of $\phi$.  This is consistent with our previous estimate.

It is even possible that we have seriously underestimated the role of
non-charmed final states in hadronic $\ppp$ decays.  If so, there is a chance
of reconciling the smaller cross section for $e^+ e^- \to \ppp$ measured by the
Mark III Collaboration using a comparison of single-charm and double-charm
production, $\sigma(\ppp) = 5.0 \pm 0.5$ nb \cite{MkIII}, with higher values
obtained by other groups using direct measurement \cite{LGW,XB,MkII,BES},
whose average I find to be $8.0 \pm 0.7$ nb.\footnote{The same average was
found in \cite{Zhu} without the data of \cite{BES}.}  This possible discrepancy
was a factor motivating the studies in Refs.\ \cite{Zhu,Walid}.  Those
and related searches need to be performed with greater sensitivity
and with off-resonance running in order to determine backgrounds from such
processes as $e^+ e^- \to \gamma^* \to {\rm charmless~hadrons}$.  In any event,
the search for the ``missing final states'' of the $\pp$ among the decay
products of the $\ppp$ is a reasonable goal of foreseen studies \cite{Wkshp}.

\section{Implications for $B$ decays}

A key observation in Ref.\ \cite{Suz01} with regard to the additional
contribution to $\pp$ hadronic decays is that it is likely to have
a large final-state phase, in order to interfere destructively with the
pertubative $3g$ contribution in the $\rho \pi$ and $K \bar K^*(892) + {\rm
c.c.}$ channels.  If this new contribution is due to rescattering into
non-charmed final states through charmed particle pairs, it is exactly
the type of contribution proposed in Refs.\ \cite{fsi,Dun,Ciu,KLS} in which
the decay $\bar b \to \bar c c \bar s$ or $\bar b \to \bar c c \bar d$
contributes to a penguin amplitude with a large strong phase.  Several
implications of this possibility were reviewed in \cite{fsi}, and others
have been pointed out in \cite{Ciu}.  These include the following:

\begin{enumerate}

\item The semileptonic branching ratio ${\cal B}(B \to X \ell \nu)$ can be
diminished with respect to the theoretical prediction if the penguin amplitude
leads to a net enhancement of $\bar b \to \bar s$ and $\bar b \to \bar d$
transitions.  The enhancement need not be large enough to conflict with
any experimental upper limits on such transitions, which are in the range
of a few percent of all $B$ decays \cite{slims}.

\item The number $n_c$ of charmed particles per average $B$ decay can be
reduced by the reannihilation of $c \bar c$ to light quarks.  The degree
to which this improves agreement with experiment is a matter of some debate
\cite{Lenz}, since a recent SLD measurement \cite{SLD} finds $n_c = 1.238
\pm 0.027 \pm 0.048 \pm 0.006$, closer to theoretical expectations than earlier
values \cite{Barker}.

\item The enhancement of the inclusive branching ratio ${\cal B}(B \to \eta'
X)$ \cite{CLEOeta} in comparison with theoretical expectations \cite{incl} can
be explained.

\item The required additional contribution \cite{eta} to the exclusive
branching ratios ${\cal B}(B \to K \eta')$ \cite{CLEOeta}, in comparison with
the penguin contribution leading to $B^0 \to K^+ \pi^-$ or $B^+ \to K^0 \pi^+$,
can be generated.

\item In any $B \to K \pi$ process in which the dominant penguin amplitude
interferes with tree-amplitude contributions, notably in $B^+ \to \pi^0
K^+$ and $B^0 \to K^+ \pi^-$, a CP-violating asymmetry can occur up to the
maximum allowed by the ratio of the tree to penguin amplitudes' magnitudes.
This asymmetry, estimated to be about 1/3 in Ref.\ \cite{fsi}, is not yet
excluded by experiment \cite{CLEOasy}.  The enhancement of the penguin
amplitude by the intrinsically non-perturbative charm rescattering mechanism
seems to fall outside the purview of the essentially perturbative approach
of Ref.\ \cite{BBNS}, so we would not expect to encounter it in that
treatment.

\end{enumerate}

The charm rescattering model for suppression of $\pp \to \rho \pi$ and related
decays has no {\it a priori necessity} for the final state phase to be
large \cite{Suz01}.  Additional evidence for such
a large final-state phase in closely related processes would be the presence of
large direct CP-violating symmetries in $B^+ \to \pi^0 K^+$ and $B^0 \to K^+
\pi^-$, with similar expected asymmetries for the two processes
\cite{Ciu,KLS,comb,MN}.  Since the process $B^+ \to \pi^+ K^0$ is not expected
to have a tree contribution, we expect it to have a much smaller
CP-violating asymmetry.  Present data \cite{CLEOasy} are consistent at the
level of 10--20\% with vanishing asymmetry for all three processes:
\beq
{\cal A}(K^+ \pi^-) = -0.04 \pm 0.16,~~
{\cal A}(K^+ \pi^0) = -0.29 \pm 0.23,~~
{\cal A}(K_S \pi^+) = 0.18 \pm 0.24.
\eeq

\section{Conclusions}

The coupling of $\pp$ and $\ppp$ to charmed particle pairs can lead to
S--D-wave mixing, the distortion of the relative branching ratios of the $\ppp$
to $\gamma + \chi_c$ final states, and the suppression of some decay modes of
$\pp$ and their appearance instead in products of the $\ppp$.  If $\ppp$ to
$\gamma + \chi_{c2}$ is observed at a branching ratio level exceeding a couple
of parts in $10^4$, this will be evidence for S--D-wave mixing, while the
branching ratio for $\ppp$ to $\gamma + \chi_{c0}$ is expected to be a percent,
give or take a factor of 2.  A similar branching ratio is expected for {\it
hadronic} charmless decays of $\ppp$.  This picture provides
a rationale for large observed $\bar b \to \bar s$ penguin amplitudes in
$B$ meson decays, and would be further supported by the observation of large
direct CP-violating asymmetries in the decays $B^+ \to \pi^0 K^+$ and
$B^0 \to K^+ \pi^-$.

\section*{Acknowledgments}

I thank San Fu Tuan for asking questions which led to this investigation and
for useful comments, and Thorsten Feldmann,
David Hitlin, Kenneth Lane, and Jon J. Thaler for
discussions.  This work was supported in part by the United
States Department of Energy through Grant No.\ DE FG02 90ER40560.

\def \ajp#1#2#3{Am.\ J. Phys.\ {\bf#1}, #2 (#3)}
\def \apny#1#2#3{Ann.\ Phys.\ (N.Y.) {\bf#1}, #2 (#3)}
\def \app#1#2#3{Acta Phys.\ Polonica {\bf#1}, #2 (#3)}
\def \arnps#1#2#3{Ann.\ Rev.\ Nucl.\ Part.\ Sci.\ {\bf#1}, #2 (#3)}
\def \b97{{\it Beauty '97}, Proceedings of the Fifth International
Workshop on $B$-Physics at Hadron Machines, Los Angeles, October 13--17,
1997, edited by P. Schlein}
\def \art{and references therein}
\def \cmts#1#2#3{Comments on Nucl.\ Part.\ Phys.\ {\bf#1}, #2 (#3)}
\def \cn{Collaboration}
\def \cp89{{\it CP Violation,} edited by C. Jarlskog (World Scientific,
Singapore, 1989)}
\def \ctp#1#2#3{Commun.\ Theor.\ Phys.\ {\bf#1}, #2 (#3)}
\def \efi{Enrico Fermi Institute Report No.\ }
\def \epjc#1#2#3{Eur.\ Phys.\ J. C {\bf#1}, #2 (#3)}
\def \f79{{\it Proceedings of the 1979 International Symposium on Lepton and
Photon Interactions at High Energies,} Fermilab, August 23-29, 1979, ed. by
T. B. W. Kirk and H. D. I. Abarbanel (Fermi National Accelerator Laboratory,
Batavia, IL, 1979}
\def \hb87{{\it Proceeding of the 1987 International Symposium on Lepton and
Photon Interactions at High Energies,} Hamburg, 1987, ed. by W. Bartel
and R. R\"uckl (Nucl.\ Phys.\ B, Proc.\ Suppl., vol.\ 3) (North-Holland,
Amsterdam, 1988)}
\def \ib{{\it ibid.}~}
\def \ibj#1#2#3{~{\bf#1}, #2 (#3)}
\def \ichep72{{\it Proceedings of the XVI International Conference on High
Energy Physics}, Chicago and Batavia, Illinois, Sept. 6 -- 13, 1972,
edited by J. D. Jackson, A. Roberts, and R. Donaldson (Fermilab, Batavia,
IL, 1972)}
\def \ijmpa#1#2#3{Int.\ J.\ Mod.\ Phys.\ A {\bf#1}, #2 (#3)}
\def \ite{{\it et al.}}
\def \jhep#1#2#3{JHEP {\bf#1}, #2 (#3)}
\def \jpb#1#2#3{J.\ Phys.\ B {\bf#1}, #2 (#3)}
\def \lg{{\it Proceedings of the XIXth International Symposium on
Lepton and Photon Interactions,} Stanford, California, August 9--14 1999,
edited by J. Jaros and M. Peskin (World Scientific, Singapore, 2000)}
\def \lkl87{{\it Selected Topics in Electroweak Interactions} (Proceedings of
the Second Lake Louise Institute on New Frontiers in Particle Physics, 15 --
21 February, 1987), edited by J. M. Cameron \ite~(World Scientific, Singapore,
1987)}
\def \kdvs#1#2#3{{Kong.\ Danske Vid.\ Selsk., Matt-fys.\ Medd.} {\bf #1},
No.\ #2 (#3)}
\def \ky85{{\it Proceedings of the International Symposium on Lepton and
Photon Interactions at High Energy,} Kyoto, Aug.~19-24, 1985, edited by M.
Konuma and K. Takahashi (Kyoto Univ., Kyoto, 1985)}
\def \mpla#1#2#3{Mod.\ Phys.\ Lett.\ A {\bf#1}, #2 (#3)}
\def \nat#1#2#3{Nature {\bf#1}, #2 (#3)}
\def \nc#1#2#3{Nuovo Cim.\ {\bf#1}, #2 (#3)}
\def \nima#1#2#3{Nucl.\ Instr.\ Meth. A {\bf#1}, #2 (#3)}
\def \np#1#2#3{Nucl.\ Phys.\ {\bf#1}, #2 (#3)}
\def \npbps#1#2#3{Nucl.\ Phys.\ B Proc.\ Suppl.\ {\bf#1}, #2 (#3)}
\def \os{XXX International Conference on High Energy Physics, Osaka, Japan,
July 27 -- August 2, 2000}
\def \PDG{Particle Data Group, D. E. Groom \ite, \epjc{15}{1}{2000}}
\def \pisma#1#2#3#4{Pis'ma Zh.\ Eksp.\ Teor.\ Fiz.\ {\bf#1}, #2 (#3) [JETP
Lett.\ {\bf#1}, #4 (#3)]}
\def \pl#1#2#3{Phys.\ Lett.\ {\bf#1}, #2 (#3)}
\def \pla#1#2#3{Phys.\ Lett.\ A {\bf#1}, #2 (#3)}
\def \plb#1#2#3{Phys.\ Lett.\ B {\bf#1}, #2 (#3)}
\def \pr#1#2#3{Phys.\ Rev.\ {\bf#1}, #2 (#3)}
\def \prc#1#2#3{Phys.\ Rev.\ C {\bf#1}, #2 (#3)}
\def \prd#1#2#3{Phys.\ Rev.\ D {\bf#1}, #2 (#3)}
\def \prl#1#2#3{Phys.\ Rev.\ Lett.\ {\bf#1}, #2 (#3)}
\def \prp#1#2#3{Phys.\ Rep.\ {\bf#1}, #2 (#3)}
\def \ptp#1#2#3{Prog.\ Theor.\ Phys.\ {\bf#1}, #2 (#3)}
\def \rmp#1#2#3{Rev.\ Mod.\ Phys.\ {\bf#1}, #2 (#3)}
\def \rp#1{~~~~~\ldots\ldots{\rm rp~}{#1}~~~~~}
\def \rpp#1#2#3{Rep.\ Prog.\ Phys.\ {\bf#1}, #2 (#3)}
\def \sing{{\it Proceedings of the 25th International Conference on High Energy
Physics, Singapore, Aug. 2--8, 1990}, edited by. K. K. Phua and Y. Yamaguchi
(Southeast Asia Physics Association, 1991)}
\def \slc87{{\it Proceedings of the Salt Lake City Meeting} (Division of
Particles and Fields, American Physical Society, Salt Lake City, Utah, 1987),
ed. by C. DeTar and J. S. Ball (World Scientific, Singapore, 1987)}
\def \slac89{{\it Proceedings of the XIVth International Symposium on
Lepton and Photon Interactions,} Stanford, California, 1989, edited by M.
Riordan (World Scientific, Singapore, 1990)}
\def \smass82{{\it Proceedings of the 1982 DPF Summer Study on Elementary
Particle Physics and Future Facilities}, Snowmass, Colorado, edited by R.
Donaldson, R. Gustafson, and F. Paige (World Scientific, Singapore, 1982)}
\def \smass90{{\it Research Directions for the Decade} (Proceedings of the
1990 Summer Study on High Energy Physics, June 25--July 13, Snowmass, Colorado),
edited by E. L. Berger (World Scientific, Singapore, 1992)}
\def \tasi{{\it Testing the Standard Model} (Proceedings of the 1990
Theoretical Advanced Study Institute in Elementary Particle Physics, Boulder,
Colorado, 3--27 June, 1990), edited by M. Cveti\v{c} and P. Langacker
(World Scientific, Singapore, 1991)}
\def \yaf#1#2#3#4{Yad.\ Fiz.\ {\bf#1}, #2 (#3) [Sov.\ J.\ Nucl.\ Phys.\
{\bf #1}, #4 (#3)]}
\def \zhetf#1#2#3#4#5#6{Zh.\ Eksp.\ Teor.\ Fiz.\ {\bf #1}, #2 (#3) [Sov.\
Phys.\ - JETP {\bf #4}, #5 (#6)]}
\def \zpc#1#2#3{Zeit.\ Phys.\ C {\bf#1}, #2 (#3)}
\def \zpd#1#2#3{Zeit.\ Phys.\ D {\bf#1}, #2 (#3)}

\end{document}